\documentclass[9pt,twocolumn,twoside]{osajnl}
\journal{ol} 
\setboolean{shortarticle}{true}
\usepackage{graphicx} 
\usepackage{subfigure} 
\usepackage{mathtools}
\usepackage{amsthm}
\usepackage{latexsym,amsmath,amssymb} 
\usepackage{tikz} 
\usepackage{caption}
\usepackage{listings}
\usepackage{color} 
\definecolor{mygreen}{RGB}{28,172,0} 
\definecolor{mylilas}{RGB}{170,55,241}

\title{\emph{Seeing the Bigger Picture:} Improving Otsu's Thresholding Method of Global Binarization Using Ring Theory for Ultrasonographies of Congestive Heart Failure}

\author[1]{Alisa Rahim}
\author[2]{Esley Torres}

\affil[1]{Liberal Studies Department, New York University (NYU), 726 Broadway, 6th Floor, New York, NY 10003.}
\affil[2]{Centro de Investigación en Ciencias (CINC), Instituto de Investigación en Ciencias Básicas y Aplicadas (IICBA), Universidad Autónoma del Estado de Morelos (UAEM), Mexico.}

\begin{abstract}
Ring Theory states that a ring is an algebraic structure where two binary operations can be performed among the elements: addition and multiplication. Binarization is a method of image processing where values within pixels are reduced to a scale from zero to one, with zero representing most absence of light and one representing most presence of light. Currently, sonograms — computer images composed by movement of sound waves — are implemented in scanning for congestive heart failure. However, the renowned ``Playboy Bunny'' symbol representing the ailment becomes increasingly difficult to isolate due to surrounding organs and lower quality image productions. This paper examines the Otsu thresholding method and incorporates new elements to account for different image features meant to better isolate congestive heart failure indicators in ultrasound images. 
\end{abstract}

\begin{document}

\maketitle

\section{Introduction}
\setlength\parindent{12pt}
\text{}

Ring Theory is a branch of abstract algebra within the field of pure mathematics, stating that a ring is a set of elements with two binary operations — addition and multiplication. The term ”elements” is used rather than ”numbers” because rings are used to generalize complex mathematical concepts, including matrices and polynomials with real coefficients. These operations, along with subtraction, can be performed within a ring. However, division may not always be possible among the elements in a ring, and multiplication is not guaranteed to be commutative among all rings. To subtract within a ring would essentially mean to add an element to its additive inverse. Likewise, to divide would mean to multiply an element by its multiplicative inverse. Doing so would bring the element back to the additive identity zero and the multiplicative identity one, respectively \cite{rahim_2020}.

Binarization is a subprocess of image segmentation whereby an image becomes a binary resolution - one that is grayscale and where the pixel values range from zero to one in regards to presence of darkness. The process of binarization starts with the image being converted to grayscale, then applying an adaptive threshold to the final resolution.  The adaptive threshold allows for the image to output different iterations of the same picture. There are two types of binarization: global (focusing on the entire image) and local (honing in on a specific region). Binarization, like other methods of image segmentation, has various uses — including document scanning, forensic analysis, and medical imagery \cite{zhu_Convolutional-Network_2020}.

Congestive heart failure is an illness where the heart cannot control blood flow adequately and often requires conduction of medical imagery — including echocardiograms (EKGs), magnetic resonance imaging (MRIs), or computed tomography (CT) scans among other methods. The heart may fail to pump blood (systolic heart failure) or fill up with blood (diastolic heart failure) in an efficient manner\cite{bayraktar_hepatic_2007}. Previous research acknowledges that congestive heart failure is detected by a “Playboy Bunny” sign: denoted by heavy demarcations in the left, middle, and right hepatic veins, located by the liver and within the inferior vena cava. A medical image production can be inaccurate due to either the presence of surrounding organs or underperforming technology, which can lead to an incorrect or missed diagnosis altogether. This research seeks to improve the detection of congestive heart failure by revising Otsu’s thresholding method, a means of global binarization.

\begin{figure}[H]
\centering
\begin{subfigure}
  \centering
  \includegraphics[scale=.405]{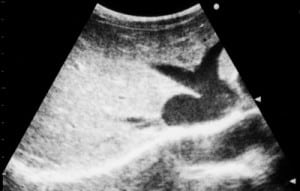}
  \label{subfig:bunny1_intro}
\end{subfigure}
\begin{subfigure}
  \centering
  \includegraphics[scale=.34]{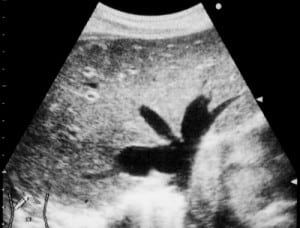}
  \label{subfig:bunny2_intro}
\end{subfigure}
\caption{Examples of ultrasonographies showing the "Playboy Bunny" symbol of congestive heart failure. }
\label{fig:bunny_intro}
\end{figure}

\section{Related Literature}
Image segmentation is a popular practice in the field of medical imagery. Ritu and Ginni Garg assert the following in their publication entitled “Brain Tumor Detection and Classification based on Hybrid Ensemble Classifier”: “The challenging task in Brain Tumor is due to high variability and inherent MRI data characteristics, e.g., variability in tumor sizes or shapes, tumor detection, area calculation, segmentation, classification, and finding uncertainty in segmented region. The most significant task in image understanding is image segmentation because it helps in feature extraction, area calculation, and significance in many real-life applications. It can be used, for example, estimation of tumor volume, tissue classification, blood cell delineation, and localization of tumors, matching of an atlas, surgical planning, and image registration” \cite{garg_brain_2021}. Garg reiterates difficulties posed in medical image segmentation, applicable to imagery any bodily organ. Current thresholding methods are still vulnerable to inciting misdiagnoses due to poorly produced scans. 

Sensitivities in medical imagery apparatus can also lead to faulty image reproductions. In “Long-evolution ascites in a patient with constrictive pericarditis”, Nunes and colleagues wrote: “The electrocardiogram is not characteristic and may reveal nonspecific repolarization changes and low-voltage electrical activity. The presence of pericardial calcification on chest radiograph is typical but not routinely found. Echocardiography offers important clues that strongly suggest the diagnosis but CT scan and MRI are the most sensitive methods to identify pericardial thickening… CP is a complex disease, systemic in nature, and many of the diagnostic problems come from its insidious course and lack of distinctive cardiopulmonary symptoms. A particular subgroup of patients may represent an interesting diagnostic challenge, presenting with ascites and signs of chronic liver disease.” (Nunes et al., 2016) A single heart scan can bring back issues with surrounding organs, or images that reflect them. For instance, signs of congestive heart failure could be intertwined with symptoms being shown within the liver. Failures in deduction can lead to an incorrect diagnosis or medicalization of a different body system \cite{garg_brain_2021}.

Additionally, different illnesses show similar symptoms, and an objective heart scan will not be sufficient enough in identifying the exact ailment in question. Yusuf Bayraktar, MD states in “Hepatic venous outflow obstruction: Three similar syndromes” that “although patients with [hepatic venous outflow obstruction] generally present with abdominal pain due to hepatomegaly, and jaundice and ascites due to portal hypertension; chronic [Budd-Chiari Syndrome] patients may first present with cirrhosis and its complications. BCS patients with inferior vena cava obstruction may also have leg edema and venous collaterals over the trunk. Additionally, signs and symptoms of heart failure such as jugular venous distention, leg edema, and dyspnea may be seen in patients with [congestive hepatopathy]. Nonetheless, the histological findings in all three syndromes are almost identical and include sinusoidal congestion and hepatocyte necrosis predominating in perivenular areas of hepatic acini which eventually leads to bridging fibrosis between adjacent central veins… The laboratory findings in [veno-occulsive disease], BCS, and CH are also very similar.” (Bayraktar 2007) Different heart conditions can coexist, which can lead doctors or medical technicians to improperly diagnose a patient. Ultrasounds of a heart can show passive liver congestion, inferior vena cava probes, or enlarged lymph nodes. 

Implementing or improving on current thresholding methods is essential to the continuation of better CT, MRI, and/or PET scans among others. In “Automatic image thresholding using Otsu’s method and entropy weighting scheme for surface defect detection”, Truong et al. dessert the following about the Otsu thresholding method: “Otsu’s method, a state-of-the-art automatic thresholding technique, is the basis for several automatic defect detection proposals. It determines optimal threshold values that maximize the between-class variances of the foreground and background. Studies demonstrate that Otsu’s method is effective for thresholding a histogram with bimodal or multimodal distribution. Otsu’s method fails if the histogram is unimodal or almost unimodal (Ng, 2006). Therefore, it provides acceptable results for thresholding general real-life images and fails when being applied in visual inspection systems because of the nature of the input images” \cite{truong_Otsu_2018}. When an image has a bi-modal histogram, its foreground and background pixels are easily discernible. This is not always the case with medical image scans, especially with ailments that continue developing and spreading across the body in stages (cancers, tumors, etc). At said ailments’ final stages, images can come back and appear unimodal. Therefore, examining and revising the Otsu thresholding method is vital to the production of higher quality scans.

\section{Theoretical Aspects}

\subsection{Ring Theory}
\setlength\parindent{24pt}
\text{}
A ring must satisfy the following axioms:
\begin{itemize}
  \item The ring, under addition, is an abelian group. 
  \item The multiplication operation is associative, and therefore closed.
  \item All operations satisfy the distributive law of multiplication over addition. 
\end{itemize}

An example of a ring includes the set of real polynomials. The set of real polynomials can be denoted as:

$$R[x]={a_n x^n+a_{n-1}x^{n-1}+\ldots +a_1x+a_0\; |\; a_i\in R}.$$

Within this ring, you can freely add, subtract, and multiply one polynomial, essentially an element within the ring, to get another polynomial - another element. The additive identity is presented as zero. Since zero is a constant polynomial, this value is also considered to be an element in the ring of real polynomials. The multiplicative identity is presented as one. Since multiplication is always commutative among all polynomials, the ring of real polynomials is deduced as a commutative ring with an identity element. 

\subsection{Otsu Thresholding Method}
The Otsu thresholding method has both mathematical and computational representations. When calculating the appropriate threshold, users must look for the minimum within class variance (sum of 2 variances multiplied by associated weights):
$$(\sigma_W^2) = W_b (\sigma_b^2) + W_f(\sigma_f^2)$$
and maximum between class variance (largest difference between overall variance and within class variance):
$$\sigma_B^2 = \sigma^2 - \sigma_W^2$$

where:
\begin{itemize}
    \item $\sigma_W^2$ represents within class variance,
    \item $W_b$ represents the weight (light presence) of the background pixels,
    \item $\sigma_b^2$ represents the variance of the background,
    \item $W_f$ represents the weight of the foreground pixels, 
    \item $\sigma_f^2$ represents the variance of the background, 
    \item $\sigma^2$ represents the overall image variance, and
    \item $\sigma_B^2$ represents the between class variance. 
\end{itemize}

\subsection{Gaussian Filtering}
Gaussian filters are used to smooth images and remove any noise \cite{gaussian_filtering_2010}, denoted by: 
$$G(x) = \frac{1}{\sqrt{2\pi\sigma^2}}\cdot \exp{\left(-\frac{x^2}{2\sigma^2}\right)},$$
\noindent where $\sigma$ represents the distribution standard deviation, with the mean assumed to be zero.

\begin{table}[htbp]
\centering
\caption{\bf Significant Values for Gaussian Filters when $\sigma$ = 1}
\begin{tabular}{cccc}
\hline
x & $0$ & $1$ & $2$\\
\hline
$G(x)$ & 0.399 & 0.242 & 0.05 \\
$G(x)/G(0)$ & 1 & 0.6 & 0.125\\
\hline
\end{tabular}
  \label{tab:shape-functions}
\end{table}

In MATLAB, we create a new variable, Iblur1, consisting of the target image I and the smoothing level (changed to an integer). 

\subsection{K-Means Clustering Methods}
The K-Means clustering method is used to group data where it is otherwise not explicitly segmented \cite{muthukrishnan_Math-KMeans_2018}. Muthukrishnan remarks that "assuming we have input data points $x_1$, $x_2$, $x_3$, $\ldots$, $x_n$ and value of $K$ (the number of clusters needed), we follow the below procedure:

\begin{enumerate}
\item Pick $K$ points as the initial centroids from the dataset, either randomly or the first $K$.
\item Find the Euclidean distance of each point in the dataset with the identified $K$ points (cluster centroids).
\item Assign each data point to the closest centroid using the distance found in the previous step.
\item Find the new centroid by taking the average of the points in each cluster group.
\item Repeat 2 to 4 for a fixed number of iteration or till the centroids don’t change.
\end{enumerate}

Centroids are used as the average values of each cluster, and Euclidean distance is calculated using: 
$$d(p,q) = \sqrt{(q_1 - p_1)^2 + (q_2 - p_2)^2}$$
where p and q represent two points in space, $p_1$ and $q_1$ represent the x-coordinates of the point, and $p_2$ and $q_2$ represent the y-coordinates of the point.

\begin{figure}[H]
\centering
\begin{subfigure}
  \centering
  \includegraphics[scale=0.5]{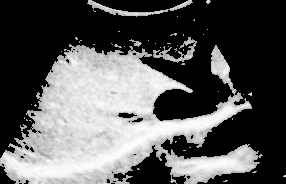}
  \label{subfig:bunny1K}
\end{subfigure}%
\begin{subfigure}
  \centering
  \includegraphics[scale=0.5]{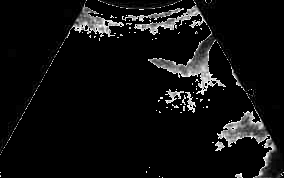}
  \label{subfig:bunny1K2}
\end{subfigure}
\begin{subfigure}
  \centering
  \includegraphics[scale=0.5]{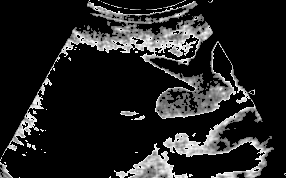}
  \label{subfig:bunny1K3}
\end{subfigure}
\caption{Image reproductions of the first "Playboy Bunny" segmentation.}
\label{fig:bunny1K}
\end{figure}

\begin{figure}[H]
\centering
\begin{subfigure}
  \centering
  \includegraphics[scale=0.5]{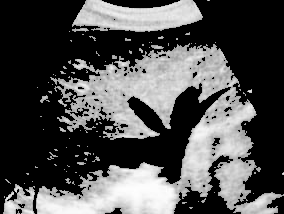}
  \label{subfig:bunny2K}
\end{subfigure}%
\begin{subfigure}
  \centering
  \includegraphics[scale=0.5]{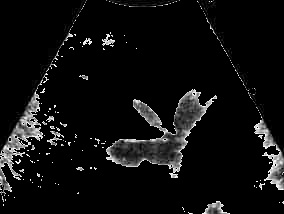}
  \label{subfig:bunny2K2}
\end{subfigure}
\begin{subfigure}
  \centering
  \includegraphics[scale=0.5]{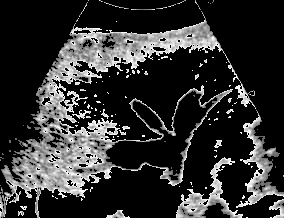}
  \label{subfig:bunny2K3}
\end{subfigure}
\caption{Image reproductions of the second "Playboy Bunny" segmentation.}
\label{fig:bunny2K}
\end{figure}

The output result is 3 different images, each with a different number of clusters and ultimately different reproduced figure. Figures \ref{fig:bunny1K} and \ref{fig:bunny2K} demonstrate $K$-Means Clustering using the images from Figure 1. 

\subsection{Ring Theory in Binarization}

For the sake of the research at hand, the method of clustering being implemented treats the pixels of an image in the same way elements of a ring are handled. In computing different segmentations of the same image, the method forms different "rings" and attempts to keep all elements within their respective clusters. The focus of this research is to reproduce an image that better indicates the renowned Playboy Bunny symbol\cite{hokama_PlayboyBS_2011}, and Ring Theory can be used to achieve this by isolating the region of interest from surrounding organs (distinguishing foreground clusters from background clusters).

\section{Improvements to Otsu Thresholding Method}

In order to improve the quality of the image reproduction, the Otsu thresholding method will instead incorporate both Gaussian filters and $K$-Means Clustering. The clustering is done first to separate the symbol pixels from the surrounding organs, and the Gaussian smoothing is encoded afterwards to remove any noise from any other elements in the environment and further isolate the region of interest. 

Since the first produced image using K-Means for both original segmentations provides the clearest distinction between the notable symbol and surrounding parts, the code runs under the assumption that $k$ = 1.

\section{Experimental Results}

\begin{figure}
\centering
  \includegraphics[width=1\linewidth]{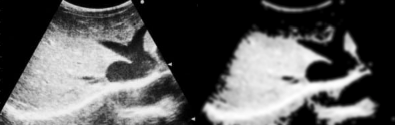}
  \caption{First original image (left) and image produced by the improved Otsu method (right).}
  \label{fig:bunny1FINAL}
\end{figure}

\begin{figure}
\centering
  \includegraphics[width=1\linewidth]{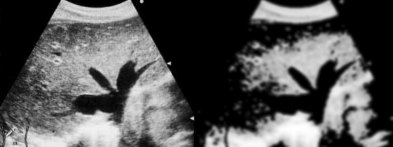}
  \caption{Second original image (left) and image produced by the improved Otsu method (right).}
  \label{fig:bunny2FINAL}
\end{figure}

Using the above code, segmentations will be run on both original images. Figures \ref{fig:bunny1FINAL} and \ref{fig:bunny2FINAL} show the original images alongside their reproductions using the improved Otsu method. 
  
\section{Conclusions}
The original method of Otsu thresholding failed to account for noise removal, which ultimately caused reproduced images to be of lower quality. The improvements made to the Otsu thresholding method herein account for both the removal of bias in clustering symbol elements from image background and smoothing the image altogether to achieve a replica of higher resolution. 

K-Means was implemented as this specific method can identify more unconventional figures, whereas other segmentation practices (i.e. Mean Shift Iterative Algorithm) are meant to identify uniform shapes or require additional training, similar to a convolutional neural network.

\section{Limitations}
\begin{itemize}
    \item This project works specifically with MATLAB, and therefore has not been tested or revised in other programming languages.
    \item Since the focus of this research was to better detect the notable “Playboy Bunny” symbol for congestive heart failure, other ailment detection methods have not been looked into, and the improvements made within this project have not been tested to better pinpoint other illness indicators. 
    \item Due to the ongoing global health crisis, access to resources and mentorship were both inevitably limited. 
    \item Image reproduction is directly correlated to the original image quality; in images with poorer production methods, isolation of the region of the interest becomes increasingly lower in quality. 
\end{itemize}

\section{Suggestions for Further Research}
\begin{itemize}
    \item Expand into other programming languages, most likely starting with Python and/or Javascript.
    \item Look into other ailments and different detection methods, most notably MRI and CT scans of brain tumors or soft tissue sarcomas. 
    \item When possible, continue networking to outside professionals and laboratories to implement improved methods into medical technology apparatus.
    \item Begin testing the new thresholding method on a larger set of images and looking into different methods of original image production. 
\end{itemize}

\medskip

\noindent\textbf{Disclosures} The authors declare no conflicts of interest.

\section{Acknowledgements}

This project would not have been made possible without the multitude of support I have received from friends, family, and fellow academic professionals. I would like to especially thank Dr. Rishi Nath of York College for first introducing me to Ring Theory. Additionally, I have both Professors Yasel Garcés Suárez and Esley Torres of the Institute of Cybernetics to thank for introducing me to binarization and assisting with the format of this paper, respectively. 

\bibliography{sample}

\setlength\parindent{24pt}
\text{}

    
    
\end{document}